\documentclass [apl,reprint,twocolumn,graphicx]{revtex4-1} 
\usepackage{amsmath}
\usepackage{color}
\usepackage{hyperref}
\hypersetup{
    colorlinks=true,
  citecolor=blue,
  linkcolor=red,
}
\usepackage{graphicx}
\usepackage{epstopdf}
\usepackage{array}
\usepackage{multirow}
\begin{document}
\title{Depth resolved chemical speciation in a superlattice structure}
\author{Gangadhar Das$^{1,2}$, A. G. Karydas$^{3,4}$, Haranath Ghosh$^{1,2}$, M. Czyzycki$^{3,5}$, A. Migliori$^{3}$, A. K. Sinha$^{1,2}$ and M. K. Tiwari$^{1,2,*}$}
\affiliation {$^1$Indus Synchrotrons Utilization Division, Raja Ramanna Centre for Advanced Technology, Indore-452013, M. P., India. \\ 
$^2$Homi Bhabha National Institute, Anushaktinagar, Mumbai-400094,India.\\
$^3$Nuclear Science and Instrumentation Laboratory, International Atomic Energy Agency, Vienna – Austria.\\
$^4$National Center for Scientific Research - Demokritos,Institute of Nuclear and Particle Physics, Athens, Aghia Paraskevi 153 10, Greece.\\
$^5$AGH University of Science and Technology, Al. A. Mickiewicza 30, 30-059 Krakow, Poland.}

\date{\today}
\begin {abstract}
We report results of simultaneous x-ray reflectivity and grazing incidence x-ray fluorescence measurements in combination with x-ray standing wave assisted depth resolved near edge x-ray absorption measurements to reveal new insights on chemical speciation of W in a W-B$_4$C superlattice structure. Interestingly, our results show existence of various unusual electronic states for the W atoms especially those sitting at the surface and interface boundary of a thin film medium as compared to that of the bulk. These observations are found to be consistent with the results obtained using first principles calculations. Unlike the conventional x-ray absorption measurements the present approach has an advantage that it permits determination of depth resolved chemical nature of an element in the thin layered materials at atomic length scale resolution.
\end{abstract}

\maketitle
\section{Introduction}
\hspace*{6pt}
Electronic states at the surface and interface boundary of a thin film medium significantly differ from that of the bulk due to strong orbital reconstruction or hybridization of near surface atoms \cite{JC7}. Such a hybridization extensively modulates the physical properties of a thin film medium. Despite the recent advances in thin film growth methodology the interface boundary between two materials in a thin film can not be realised distinctly due to atomic migration \cite{NN6}. Surface and interface alteration due to atomic scale modulation of the orbital occupation is an active area of research to unravel insights on many correlated and uncorrelated physical phenomena, such as, electronic and magneto transport properties, interlayer exchange coupling properties of various thin layered materials in condensed matter physics research \cite{EB1,ADR8,GJ8,AT8,CA9,MS9,SB7}. In order to achieve a detailed understanding about the electronic mechanism responsible for the observed physical properties in thin layered materials it is often necessary to investigate and compare unusual behaviour of orbital electrons in surfaces, interfaces and bulk \cite{JC7,EB1,MZ5,JM0,TO1}. Probing surfaces and interfaces in a superlattice structure \cite{GAPL6,JRC5} can lead to new opportunity to reveal unusual physical properties associated with a thin layered material. Despite immense interest in this direction, no appropriate technique is available that can provide depth resolved localised chemical information about a thin layered material. The conventional absorption techniques suffer from the inherent technical limitations as they do not provide depth resolved information about physical and chemical properties of nanostructured materials and their localization with respect to the substrate surface. The recent advances in the field of surface-interface analysis of thin layered materials have spurred new experimental efforts to develop methods that are far more accurate as compared to that of the conventional absorption methods.\\
\hspace*{6pt}
Fluorescence assisted x-ray standing wave (XSW) technique has become the workhorse for the surface-interface aspects of condensed matter research. As a nondestructive probe, the XSW induced fluorescence measurements at grazing incidence angles offer atomic scale depth resolution inside a thin film medium \cite{MPRB9,JZ3,YK5,DCM0,DCM9,BP8,BP3,KS7}. Recent advancements in various combined x-ray spectroscopy approaches have made it possible to obtain depth resolved chemical information on surfaces and interfaces in a nondestructive manner. In our previous work we have shown that x-ray reflectivity (XRR) together with grazing incidence x-ray fluorescence (GIXRF) measurements can be used as a sensitive probe to evaluate depth resolved microstructural parameters of a buried layer inside a W-B$_4$C superlattice structure \cite {GAPL6}. It has been demonstrated that a slight diffusion of W into a B$_4$C layer, significantly increases the density of the B$_4$C medium. Tungsten has an outer shell electronic configuration 5d$^4$ 6s$^2$ 6p$^0$. Its 5d band remains partially filled in case of divalent, trivalent, tetravalent, pentavalent compounds and completely empty in the case of the hexavalent compounds. Since the 5d, 6s and 6p electronic levels are very close to each other, hybridization of these orbitals are energetically favourable in many tungsten compounds. Such hybridization significantly modulates the nature of these higher shell energy levels \cite {EO6,UJ4,FJG5,VL7,SAG6}. Hence, depth resolved chemical speciation is often desirable to understand the underlying degradation pathways (\textit{if any}) and its influence on the interfacial chemical changes in the case of a superlattice structure.\\
\hspace*{6pt}
Here, we report depth resolved chemical speciation of a W-B$_4$C superlattice structure using combined XRR and GIXRF measurements along with the XSW assisted depth resolved x-ray absorption near edge structure (XANES) measurements. The depth selectivity of the XSW wave field inside the multilayer structure was controlled by precisely tuning the grazing incidence angle of the impinging x-ray wave field. We have carried out x-ray reflectometry measurements to investigate the anomalous behaviour of the optical constants ($\delta,\beta$) of W in the vicinity of W-L$_3$ characteristics absorption edge. Such measurements are highly useful for obtaining quantitative information on stoichiometric chemical nature and electronic distribution of the probed atom in the material. XSW assisted depth resolved XANES measurements near the W-L$_3$ absorption edge energy (E$_o$ $\sim$ 10207 eV) are carried out for two sets of W-B$_4$C superlattice structures comprising of 10 and 15 bilayer repetitions, respectively. Our results show existence of unusual electronic states for the W atoms that are present at interface boundary as compared those present in the bulk thin film medium in case of the W-B$_4$C superlattice structure. The grazing incidence x-ray diffraction (GIXRD) measurements performed on the same multilayer structure supports results of the XSW investigations. We have also carried out detailed calculations for density of states and XANES spectra for the bulk crystalline W using first principles density functional theory. Its formal similarity with the measured XANES data strengthens our findings.
\section{Experimental}
\hspace*{6pt}
W-B$_4$C periodic multilayer structures, comprising of N=15 and N=10 bilayers, used in the present work were prepared on a polished Si(100) substrate at room temperature using a DC magnetron sputtering system \cite{DPG}. The multilayer sample was deposited using Argon, as a sputtering gas medium at constant pressure of $\sim$ 5$\times$10$^{-3}$ mbar, whereas the base vacuum of the chamber was maintained at $\sim$ 2$\times$10$^{-8}$ mbar before the start of the deposition process. The combined x-ray reflectivity and grazing incidence x-ray fluorescence measurements for W-B$_4$C multilayer consisting of 15 bilayer repetitions were carried out at the x-ray reflectometer station of BL-16 beamline of Indus-2 synchrotron facility at incident x-rays energy of 10230 eV, monochromatized using a Si (111) double crystal monochromator \cite{GRSI5}. The XSW assisted depth resolved XANES measurements at W-L$_3$ absorption edge energy (E$_o$ $\sim$ 10207 eV) for W-B$_4$C multilayer of 15 bilayer repetitions was performed at the same reflectometer station. An x-ray beam of size 100 $\mu$m(v) $\times$ 10 mm(h), generated using a crossed-slits aperture was allowed for the GIXRF-XANES measurements. To measure specularly reflected x-rays from a sample reflector, an Avalanche Photo Diode (APD) detector was used. The APD detector was placed $\sim $350 mm away from the sample position. The low noise and high dynamic range of the APD detector allow us to record XRR pattern with a dynamic range up to $\sim$7 orders. A vortex spectroscopy detector (SDD) comprising an active surface area of 50 mm$^2$ and having an energy resolution of $\sim $140 eV at 5.9 keV (Mn K$\alpha$ x-rays) was placed in the plane of sample substrate at a distance 25 mm \cite{GJAAS4}. Fluorescence x-rays were measured through a Al pinhole collimator of diameter $\sim$ 1 mm to maintain a constant solid angle of the SDD detector on the sample surface \cite{WL2}. The details about the BL-16 reflectometer station are described elsewhere \cite{GRSI5,MJSR3}.  We have also carried out the XSW-XANES measurements for a W-B$_4$C multilayer consisting of N=10 bilayer repetitions at the International Atomic Energy Agency (IAEA) GIXRF-XRR experimental facility operated at the XRF beamline of Elettra Sincrotrone Trieste (BL-10.1L). GIXRD measurement was carried out at angle dispersive x-ray diffraction (ADXRD) beamline (BL-12) of Indus-2 synchrotron facility at incident x-rays energy of 15.5 keV \cite{GS6}.\\
\section{Results and Discussions}
\begin{figure*}
\includegraphics[width=0.25 \textwidth]{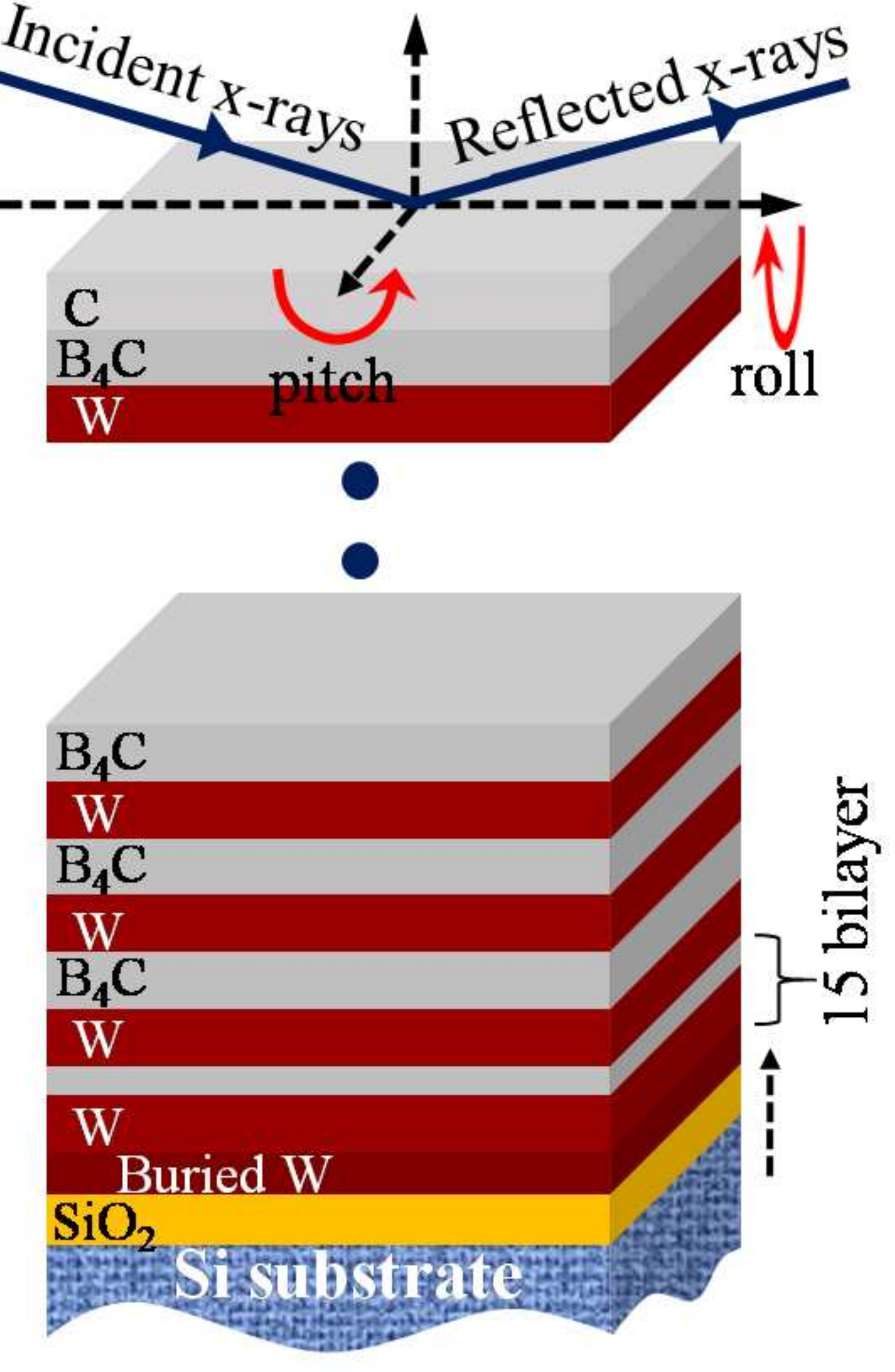}
\caption{A schematic representation showing the x-ray reflection from a W-B$_4$C superlattice structure consisting of N=15 bilayer repetitions. In this figure we have also depicted a buried interface layer of W inside the multilayer medium, on top of a Si substrate.
\label{MLS}}
 \end{figure*}

\begin{figure*}
\includegraphics[width=0.45\textwidth]{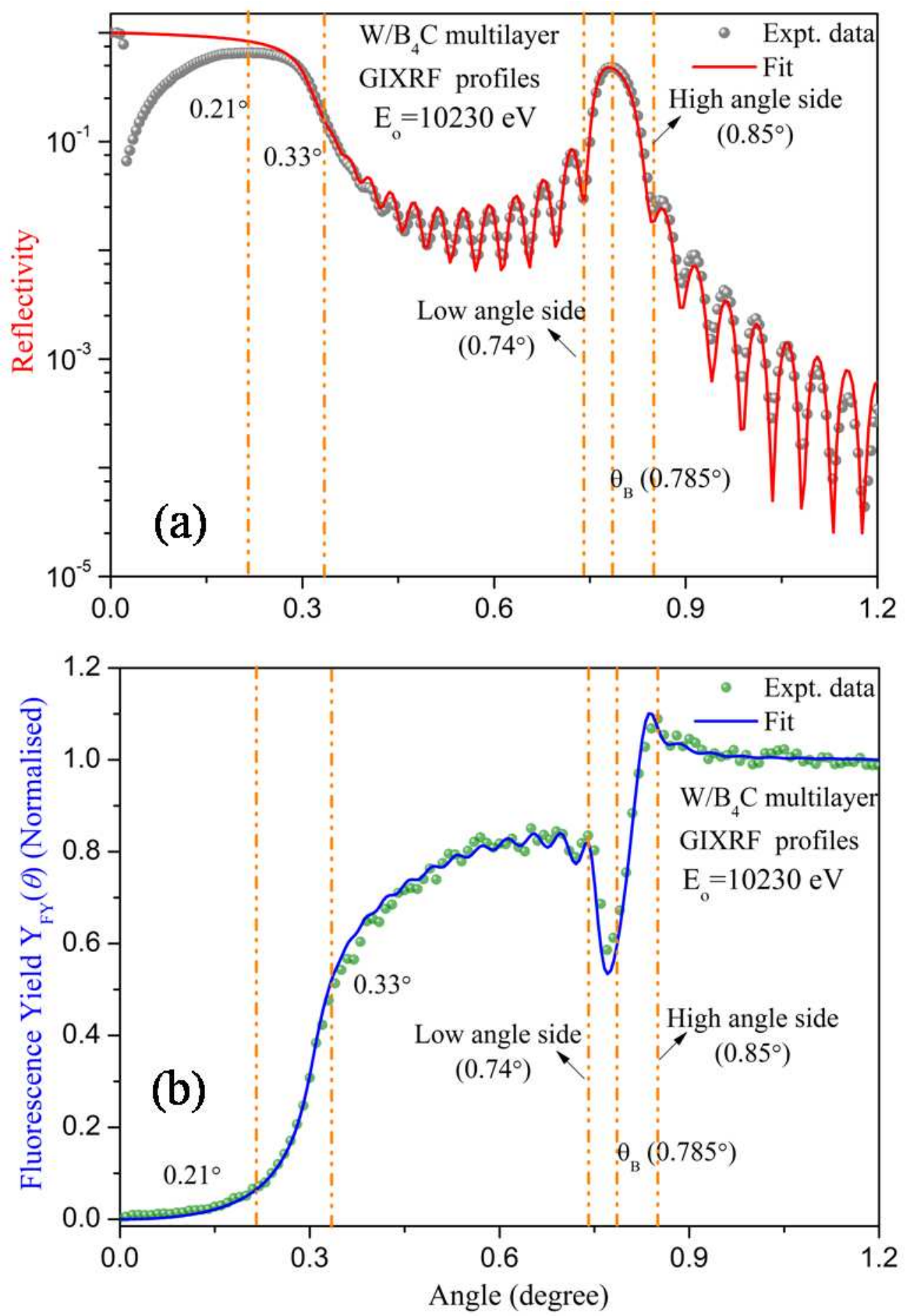}
\caption{Measured and fitted XRR and W-L$\alpha$ fluorescence profiles for W-B$_4$C multilayer structure comprising of N=15 bilayer repetitions at incident x-ray energy of 10230 eV. (a) XRR profile, and (b) W-L$\alpha$ fluorescence profile. 
\label{COML3}}
\end{figure*}

\begin{figure*}
\includegraphics[width=0.6\textwidth]{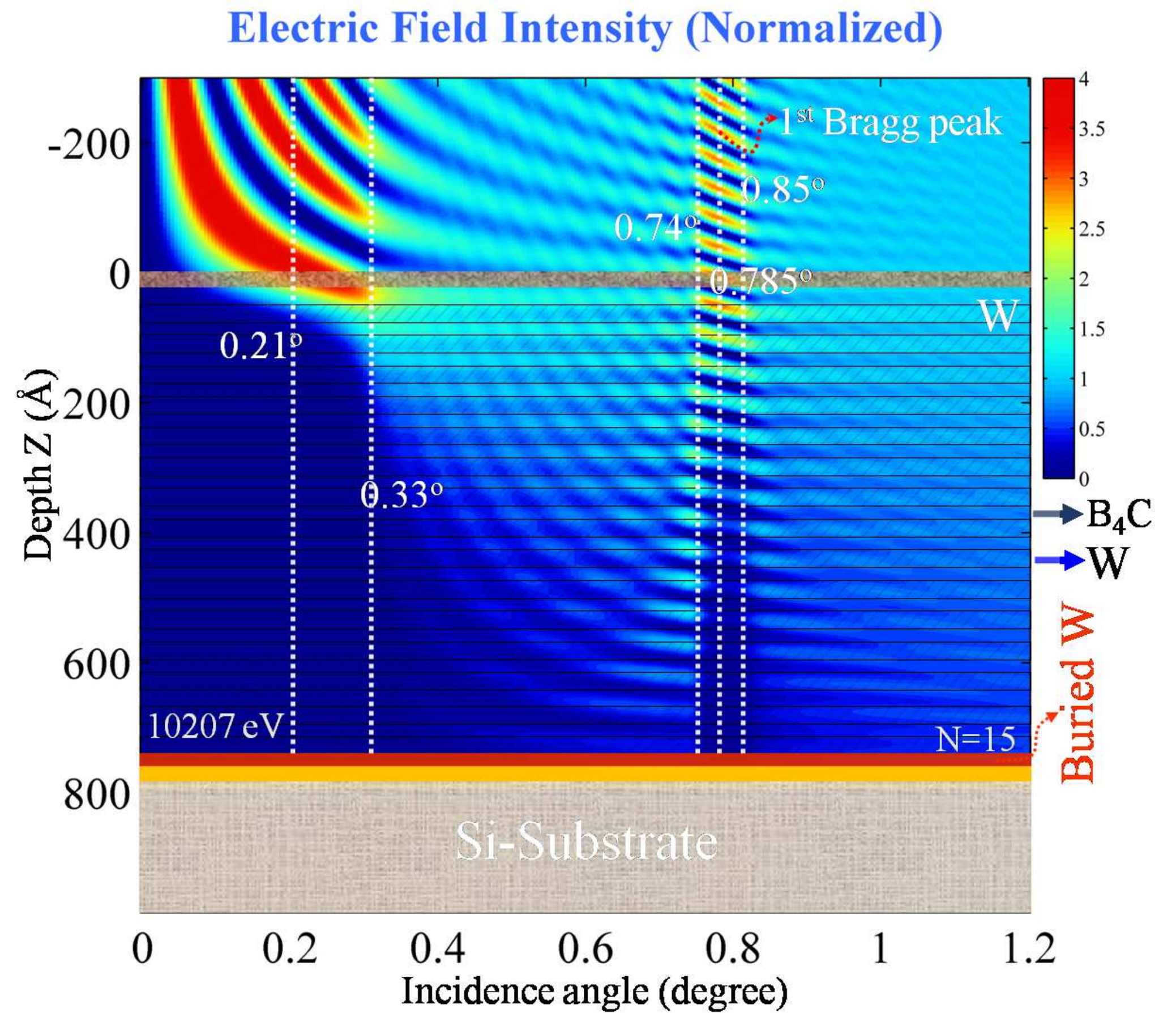}
\caption{X-ray field intensity (EFI) distribution inside a W-B$_4$C periodic multilayer structure computed using best fitted parameters obtained from the XRR and W-L$\alpha$  fluorescence measurements at incident x-ray energy of E= 10230 eV. In this figure, we have also  marked various dotted vertical lines showing different incident angles that have been opted for depth selective XANES measurements of W-B$_4$C multilayer structure.
\label{EFL3}}
\end{figure*}

\begin{figure*}
\includegraphics[width=0.55\textwidth]{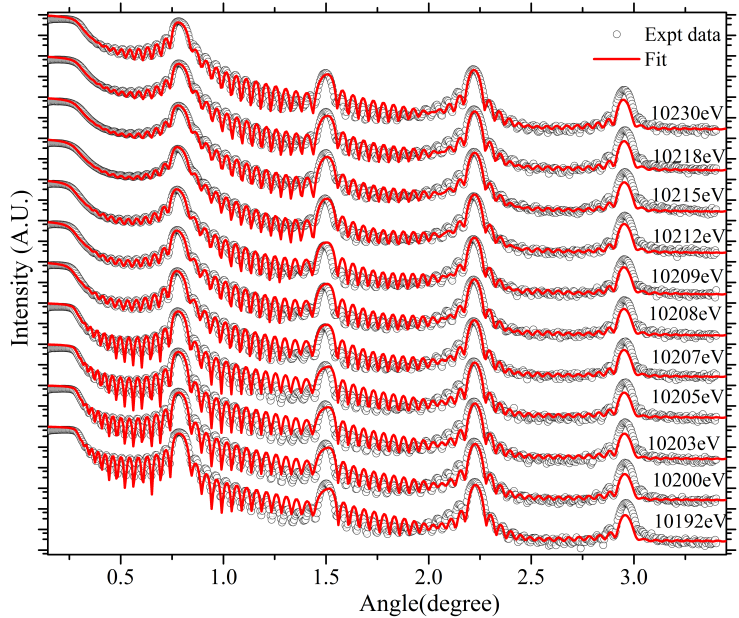}
\caption{Measured and fitted XRR profiles for W-B$_4$C multilayer structure comprising of N=15 bilayer repetitions in the incident x-ray energies of 10192-10230 eV.
\label{XRMBL3}}
\end{figure*}

\begin{figure*}
\includegraphics[width=0.45\textwidth]{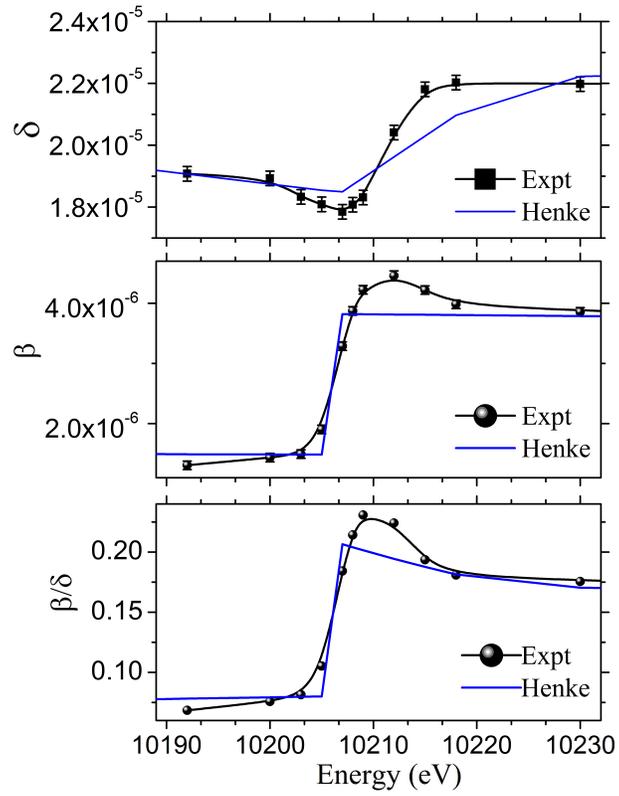}
\caption{Determined optical constants (a) $\delta$, (b) $\beta$, and (c) the ratio $\beta /\delta$ of W from the x-ray reflectivity measurements in a W-B$_4$C multilayer structure comprising of N=15 bilayer repetitions in x-ray energy region of 10192 -10230 eV. The values of optical constants calculated from the Henke table are also given for comparison.
\label{OCML}}
\end{figure*}

\begin{figure*}
\includegraphics[width=0.93\textwidth]{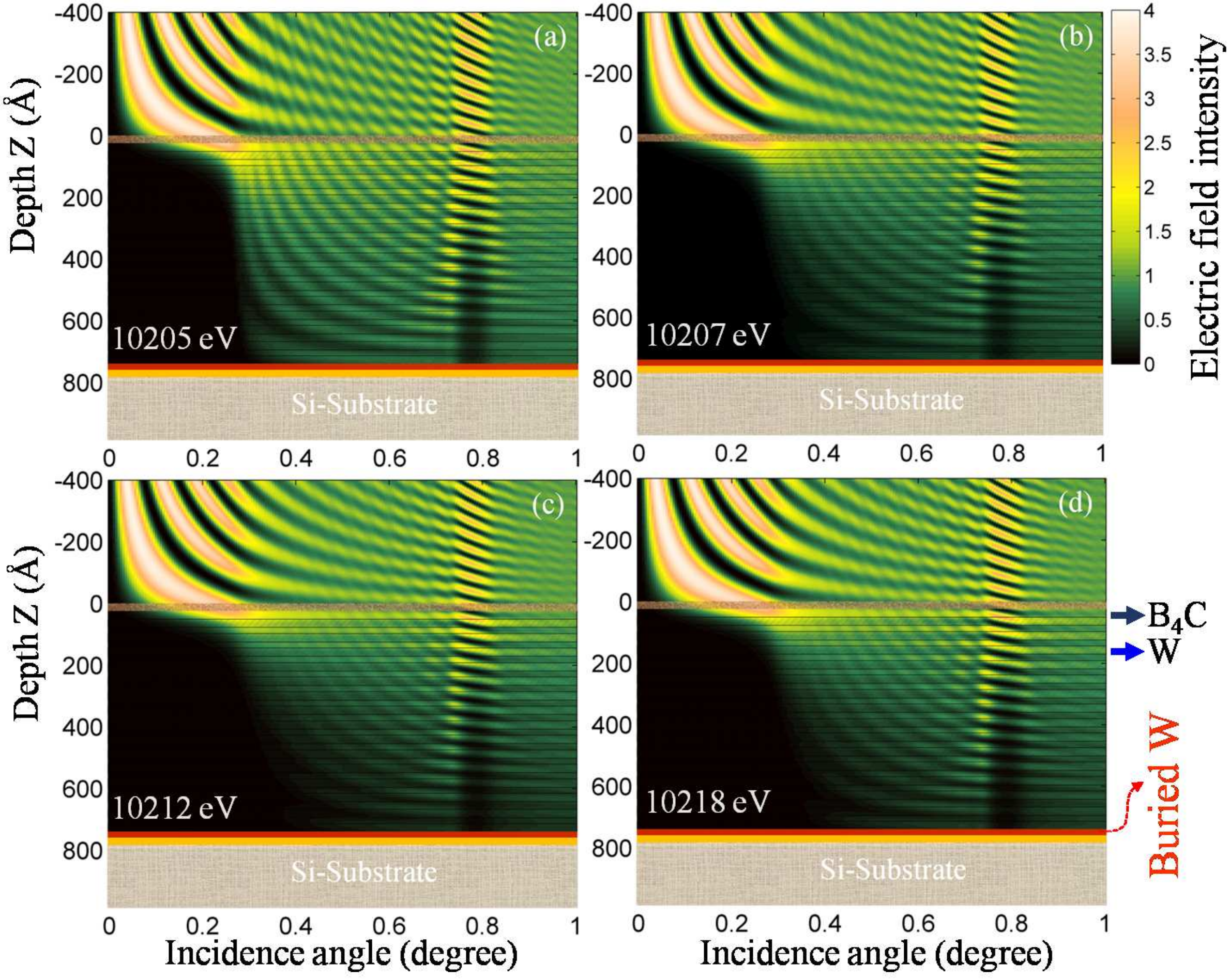}
\caption{X-ray field intensity (EFI) distribution inside a W-B$_4$C periodic multilayer structure computed using best fitted parameters obtained from the XRR and W-L$\alpha$ fluorescence measurements and determined optical constants ($\delta,\beta$) at incident x-ray energy of (a)10205 eV, (b) 10207 eV, (c) 10212 eV and (d) 10218 eV. One clearly observed the proving depth volume of XSW wave field inside W-B$_4$C superlattice structure across the W-L$_3$ absorption edge energy preserves a more or less constant volume, except a angular shift $ \theta \sim0.01^o$ at high energy side ($E \sim 10230 eV$) of the W-L$_3$ edge.
\label{EFMLV}}
\end{figure*}

\begin{figure*}
\includegraphics[width=0.98\textwidth]{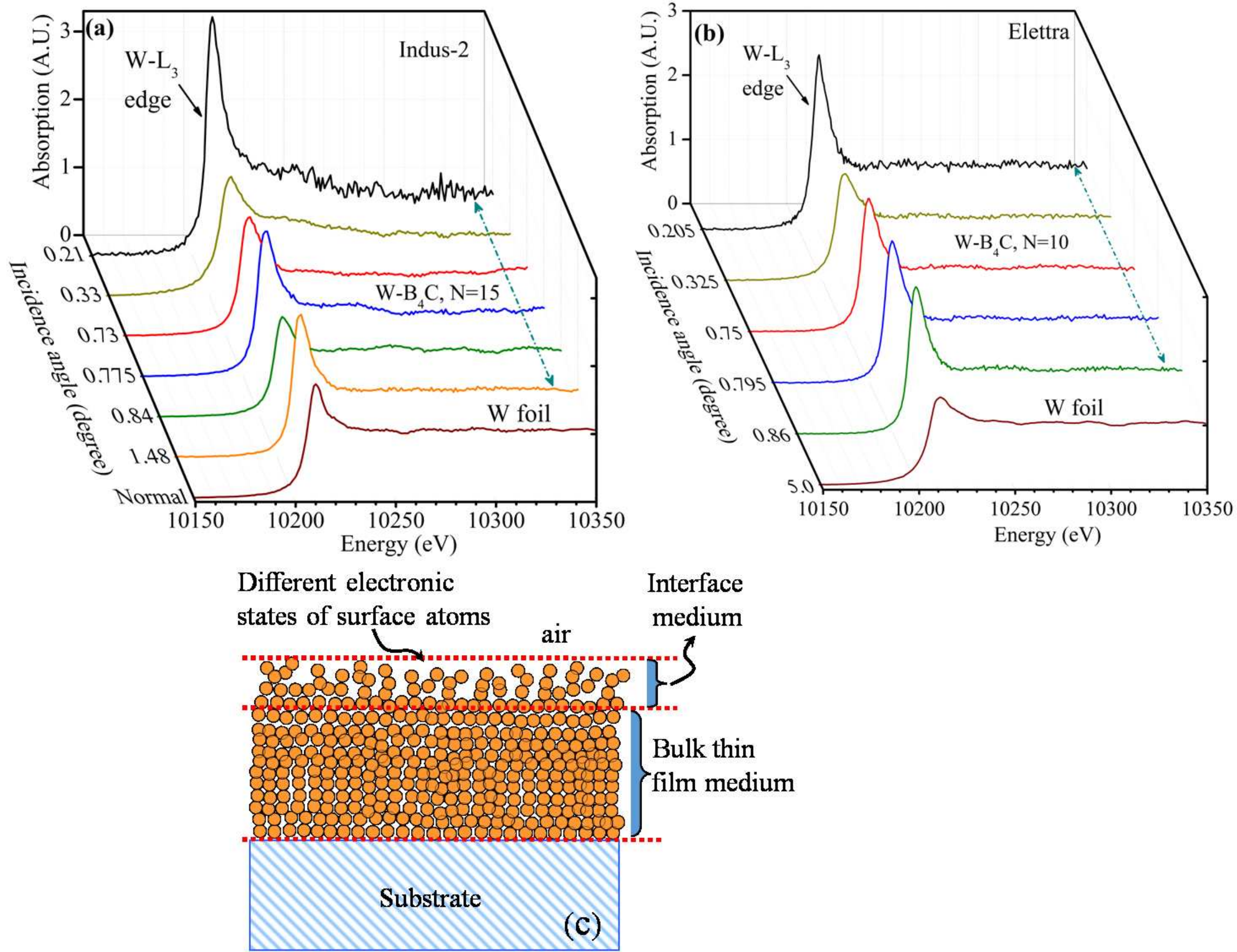}
\caption{X-ray standing wave induced XANES profiles of the W-B$_4$C multilayer structures measured at L$_3$ edge of W. (a) Measured XANES profile of a multilayer structure consisting of N=15 bilayer repetitions using BL-16 beamline of Indus synchrotron facility, and (b) W-B$_4$C multilayer structure with N=10 bilayer repetitions measured at the International Atomic Energy Agency (IAEA) GIXRF-XRR experimental facility operated at the XRF beamline of Elettra Sincrotrone Trieste (BL-10.1L). On the right hand side, (c) a schematic illustration depicts the real space distribution of W atoms at surface-interface boundary as well as inside bulk thin film medium.
\label{XANES}}
\end{figure*}

\begin{figure*}
\includegraphics[width=0.5\textwidth]{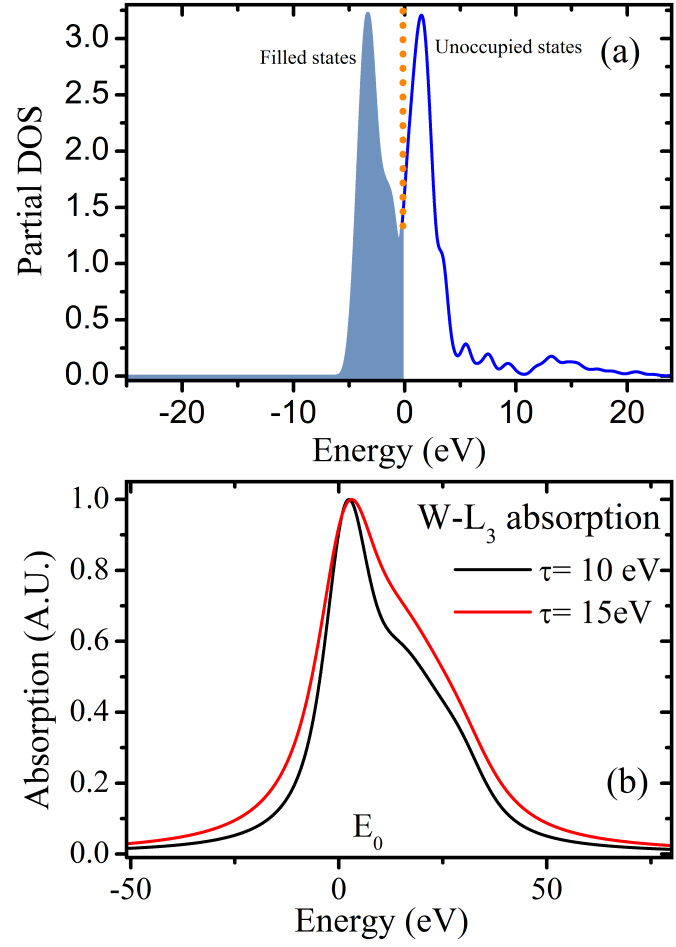}
\caption{First principles simulated (a) partial density of states of W (5d) and, (b)core level (L$_3$) absorption spectra of pure bulk bcc W.
\label{THE}}
\end{figure*}

\begin{figure*}
\includegraphics[width=0.5\textwidth]{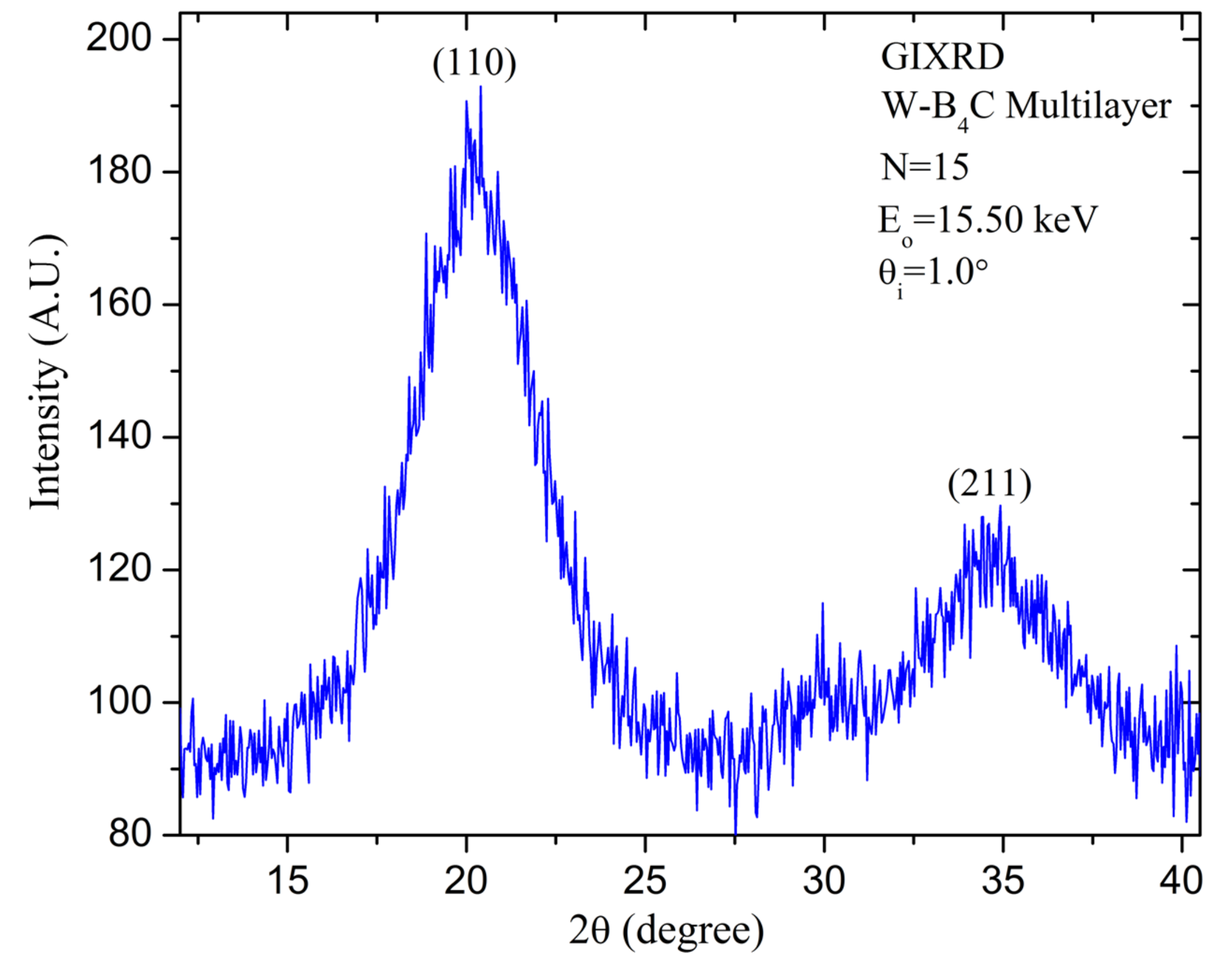}
\caption{Measured GIXRD pattern of W-B$_4$C multilayer consisting of N=15 bilayer repetitions at incident x-ray energy of 15.5 keV.  During the measurements the incidence angle of the primary x-ray beam was fixed at 1.0$^o$.
\label{GIXRDML}}
\end{figure*}

\hspace*{6pt}
Figures \ref{COML3}(a) and \ref{COML3}(b) respectively show the measured and fitted XRR and GIXRF profiles in the vicinity of the 1$^{st}$ Bragg peak for the W-B$_4$C multilayer structure consisting of 15 bilayer repetitions at 10230 eV incident x-ray energy (above the W-L$_3$ absorption edge energy E$_o$ $\sim$ 10207 eV). Simultaneous fitting of the measured XRR and GIXRF data were carried out using the CATGIXRF program \cite{MXP6}. During the model fitting  of the experimental XRR and GIXRF profiles we have considered a thin buried layer of W of thickness 16.6 \AA,  inside the multilayer structure. A schematic illustration showing the location of the W buried layer is given in Fig. \ref{MLS}. The detailed microstructural analysis of the W-B$_4$C multilayer structure is described elsewhere \cite{GAPL6}. Figure \ref{EFL3} depicts computed electric field intensity (EFI) distribution at 10230 eV x-ray energy inside the W-B$_4$C periodic multilayer structure as a function of incidence angle and film depth (z). The EFI was computed taking into account the microstructural parameters determined from the best fit results obtained using combined XRR and GIXRF analysis. From Fig. \ref{EFL3}, it can be observed that the positions of XSW antinodes vary inside the multilayer medium as grazing incidence angles of the primary x-ray beam are changed. The observed periodic Kiessig interference fringes in the XRR and GIXRF profiles are anticorrelated in nature. It can be seen from the Fig. \ref{COML3}(b) that the W-L$\alpha$ fluorescence intensity profile shows a strong XSW modulation especially at the low and high angle sides of the Bragg region. The W-L$\alpha$ fluorescence intensity is minimum at very low grazing incidence angles and starts increasing rapidly as the grazing incidence angle just crosses the critical angle ($\theta_c \approx 0.29^0$) of the film medium. This variation of W-L$\alpha$ fluorescence intensity mainly arises due to the movement of  XSW wave field in the thin film structure. At very low grazing incidence angles or below the critical angle ($\theta_c \approx 0.29^0$), there is no appreciable x-ray intensity inside the multilayer medium, but it exists only on top of the multilayer surface in the form of XSW fringes due to strong reflection (total external reflection region) from the sample surface. So, a very few W atoms, situated 2-5 nm below the top surface of superlattice, is excited by the evanescent wave and the corresponding fluorescence intensity is very weak. As the incidence angle advances through the critical angle boundary, the x-ray field starts to penetrate in different layers of the multilayer medium and consequently W-L$\alpha$ fluorescence intensity increases very rapidly. At the Bragg angle ($\theta_{Bragg} \approx 0.785^0$), an XSW field of periodicity equal to the multilayer period ($\Delta = d $) is setup inside the multilayer medium. At the low angle side ($\theta_{Low} \approx 0.74^0$) of the Bragg peak, the antinodes of the XSW field remain in the low Z layer (B$_4$C layer). As the incidence angle advances across the Bragg region, these antinodes move towards the high Z layer (W layer). However, at the Bragg peak the antinodes stand exactly at the interfaces of the low Z and high Z layer. At the high angle side ($\theta_{High} \approx 0.85^0$) of the Bragg peak, XSW antinodes completely coincide with the position of the high Z layers (W layer). Because of this movement of the  XSW antinodes intensity, the W-L$\alpha$ fluorescence intensity is strongly modulated over the Bragg region. As a result, at the high angle side of the Bragg peak, we obtained relatively higher W-L$\alpha$ fluorescence yield as compared to that of the low angle side. At higher incidence angles (after the 1$^{st}$ Bragg peak) the W-L$\alpha$ fluorescence yield more or less remains constant because the W atom is excited only with the direct x-ray beam as there is no formation of XSW field due to very weak reflected x-ray beam at high incidence angles. One may expect a light XSW modulation of the W-L$\alpha$ fluorescence yield near the 2$^{nd}$ Bragg peak. These control depth selectivity of XSW field in combination with x-ray absorption fine structure analysis provides a possibility to determine the depth selective chemical speciation of a thin film medium. This can be achieved by measuring the XSW assisted fluorescence intensity from a thin layered medium across the characteristic absorption edge of a layered material. We have selected various grazing incidence angles for the XSW assisted depth resolved x-ray absorption measurements near the W-L$_3$ absorption edge energy (E$_o$ $\sim $ 10207 eV). These angles are marked by the dotted vertical lines in Figs. \ref{COML3} and \ref{EFL3}. It may be important to mention here that the probed depth volume estimation from XSW field is usually not constant across the absorption edge of a material due to anomalous behaviour of the  complex refractive indices or optical constants of the layered material.\\
\hspace*{6pt}
In order to investigate the anomalous behaviour of the optical constants ($\delta,\beta$) of W near the characteristics L$_3$ absorption region, we have performed x-ray reflectivity measurements of the W-B$_4$C superlattice structure near the W-L$_3$ absorption energy region. Figure \ref{XRMBL3} shows the measured and fitted specular reflectivity profiles of the W-B$_4$C superlattice structure in the energy region of 10192 eV to 10230 eV. The x-ray reflectivity data were recorded with an angular step size of $\theta \approx 0.005^0$ using reflectometer station of BL-16 beamline of Indus-2 synchrotron facility. Before XRR measurements, incident x-ray energies emitted from the BL-16 beamline were calibrated by performing absorption edge measurements of a pure W foil as well as a W thin film of thickness 270\AA. While fitting XRR profiles in the energy region of 10192 eV- 10230 eV, as an initial guess we have considered measured optical constants ($\delta,\beta$) of a pure W thin film. In addition, we fixed the microstructural parameters of the W-B$_4$C superlattice structure and allowed to vary optical constants of W in a controlled manner until one obtains best fit curves to the measured XRR data. It can be seen from Fig. \ref{XRMBL3} that the measured XRR patterns match quite well with the fitted profiles except a subtle difference at the 4$^{th}$ Bragg peak location. This deviation can be attributed to the non-linear diffused scattering background produced from the multilayer sample. The reflectivity of the 1$^{st}$ Bragg peak decreases significantly near the L$_3$ absorption edge region of W due to strong anomalous absorption effects.\\
\hspace*{6pt}
The average optical constants of W in the energy region of 10192 eV - 10230 eV have been determined experimentally by fitting the measured angle dependent specular x-ray reflectivity profiles of the W-B$_4$C superlattice structure.  Measured values of the optical constants ($\delta, \beta$)  of W are plotted in the Figs. \ref{OCML}(a) and \ref{OCML}(b) respectively. For comparison, we have also plotted Henke tabulated values of optical constants of W in Fig. \ref{OCML} considering a density of W $\sim$ 17.3 g/cm$^3$. It can be seen from Fig. \ref{OCML} that the measured optical constants ($\delta, \beta$) in the energy region of 10192 eV to 10230 eV show a good agreement with the Henke tabulated values. We have observed a maximum deviation $\sim$ 5 - $7\%$ between measured and tabulated values of $\delta$ where as in case of $\beta$ values, a maximum deviation $\sim$ 10 - $15\%$ was realized. In the case of x-ray reflectivity analysis, the measurement accuracies of the optical constants usually depend on the $\beta/\delta$ ratio for all the materials \cite{JMA4,RS7}. The optical constants can be uniquely determined in a situation when $\beta/\delta \leq 0.5$. Figure \ref {OCML}(c) shows the variation of $\beta/\delta$ ratio of W in our case for the energy region of 10192 eV to 10230 eV. At a first glance, this variation seems to be in close agreement with the Henke tabulated values. We observe a maximum variation in the $\beta/\delta$  ratio of W $\sim 0.23$. In order to investigate the dependency of probed depth volume of the XSW field at different incident x-ray energy and grazing incidence angles, we have computed EFI distribution in the vicinity of the L$_3$ absorption edge of W. Figure \ref{EFMLV} represents the computed EFI distribution at (a) 10205 eV, (b) 10207 eV, (a) 10212 eV, and (a) 10218 eV x-ray energies inside the W-B$_4$C superlattice structure as a function of incidence angle and film depth (z). The EFI was computed taking into account the fitted microstructural parameters of the superlattice structure and derived optical constants obtained from the XRR analysis at different x-ray energies. From Fig. \ref{EFMLV}, it can be observed that the probed depth of the x-ray field inside the superlattice structure remains more or less constant in case if incident x-ray energies are varying in the vicinity of W-L$_3$ absorption edge. However, the locations of XSW antinodes shift by an angle of 0.01$^0$ for x-ray energies above the L$_3$ absorption edge (E $\sim $ 10218 eV).\\
\hspace*{6pt}
Figures \ref{XANES}(a) and \ref{XANES}(b) respectively show the measured and normalised XSW assisted depth resolved  x-ray absorption near edge structure measurements for two different W-B$_4$C multilayer structures consisting of 15 and 10 bilayer repetitions at various grazing incidence angles in the vicinity of L$_3$ edge energy of tungsten. In this figure, we have also plotted measured XANES profile (in the fluorescence mode) of a pure tungsten metal foil. The main absorption peak (white line) in Fig. \ref{XANES}(a) and \ref{XANES}(b) is visible due to strong transition of 2p$_{3/2}$ electrons to the partially filled 5d levels (see Fig. \ref{THE}a). It can also be seen here that the peak intensity of the white line varies at various grazing incidence angles. This is mainly attributed to different probing depths (\textit{i.e.} extinction lengths) of the x-ray wave field inside the W-B$_4$C multilayer structure at different grazing incidence angles. The observed white line features in measured XANES profiles of W-B$_4$C multilayer structure (N=15) at different grazing incidence angles, \textit{e.g.,}  $\theta$=0.73$^o$ (at low angle side), $\theta$=0.775$^o$ (at 1$^{st}$ Bragg peak) and $\theta$=0.84$^o$ (at high angle side) correspond to different probing depths of the x-ray wave field. At these incidence angles, antinodes of XSW field exist in the low Z layers (B$_4$C), at the interface of B$_4$C-W layers and in the high Z layers (W) respectively. The white line intensity at incidence angle of  $\theta$=1.48$^o$ (at 2$^{nd}$ Bragg peak) describes a normal excitation of the W-B$_4$C multilayer structure. This is due to the fact that at such a high incidence angle XSW field does not exist inside the multilayer structure because of very weak reflection of the primary x-ray beam. In this condition, all the W layers of the multilayer are excited only by the primary x-ray beam. It can also be seen that the characteristics features of the white line at $\theta$=1.48$^o$ is similar to that of the pure W metal foil. On the other hand, a subtle shift of the edge energy position (E$_o$=10207 eV) can be observed in the measured XANES profiles at grazing incidence angles of 0.21$^o$ and 0.33$^o$. This occurs mainly due to the small probing depth of the XSW wave field inside the multilayer medium at these angles. In such conditions only few W atoms sitting at the top surface of a tungsten layer are excited.  These top sub-surface W atoms experience a highly asymmetric crystal environment, which in turn decreases their binding energy as compared to those W atoms, situated at deep inside the multilayer medium. It can also be noticed from Fig. \ref{XANES} that at lower grazing incidence angles, the peak intensity of white line is relatively larger as compared to that of the higher incidence angles. This can be explained by understanding the origin of white line that arises as a results of electron transition from W-2p$_{3/2}$ orbitals to partially filled 5d orbitals in addition to the transitions that occur from W-2p$_{3/2}$ to unoccupied localised states near 5d states. These localised states emerge because of the bulk defects as well as due to surface effects (see Fig. \ref{XANES}(c)). In case of bulk, an atom remains in a homogeneous structured environment. This causes an overlap of electron wave functions of adjacent neighbouring atoms in all 3-directions uniformly. As a result the dipolar coupling between two orbitals that leads to origin of white line intensity is distributed nearly isotropically. Whereas, in the case of surface or interface states the atomic densities may be localized to its position because of lack of bulk symmetries. This would lead to sharp dipolar transitions between core levels and unoccupied surface states. The contribution of the surface effects are usually very large in the case of a thin film structure of thickness ranging in few tens of angstrom. At grazing incidence angle, the relative contribution of surface states will be larger as compared to that of the bulk defects, which in turn enhance the peak intensity of white line in the normalised XANES spectra. We have repeated the XANES measurements for a pure W thin film (thickness $\sim$ 270 \AA) at different grazing incidence angles and arrive at similar conclusions. It confirms that the enhanced white line intensity at very low grazing incidence angles arises mainly due to the surface states and self absorption effect does not play any significant role in this particular scenario \cite{PP9,WL5}. In conventional XANES measurements it has already been shown that the peak intensity of white line remains more or less unchanged in case of different oxidation states of W \cite{EO6, UJ4,PCP6}. Fig. \ref{XANES}(b) shows the normalized XANES spectra measured at XRR-GIXRF experimental station of IAEA- Elettra beamline (BL-10.1L) for a W-B$_4$C multilayer consisting of number of bilayer periods N=10. We arrive at similar conclusion after analyzing the XSW-XANES profiles presented in Fig. \ref{XANES}(b). These XANES profiles do not provide any signature of change of chemical state of W in the W-B$_4$C multilayer structures. Our results are found to be consistent with the investigations reported by Rao \textit{et al.}, \cite{PNR3}. They have shown that W and B$_4$C in the W-B$_4C$ multilayer remain stable up to an annealing temperature of 800$^o$C. Beyond this temperature (above $\sim$1000$^o$C), B$_4$C layer starts to decompose into free C or B atoms, which then react with the W atoms to produce more stable chemical phases of the tungsten \cite{PS0}.\\
\hspace*{6pt}
In order to further confirm our findings we carried out first principles studies on pure bulk crystalline bcc W (lattice parameter 3.16 \AA, space group Im-3m). We use plain wave basis set based density function theory (using CASTEP module of Material studio 16.0 \cite{SC5}) with energy cut-off 400 eV, on the fly generated norm conserving pseudo-potential. Brillouin zone is sampled in the k-space within Monkhorst-Pack scheme and grid size for SCF calculation is chosen as 16$\times$16$\times$16 with SCF tolerance 10$^{-7}$ eV/atom. Electronic exchange correlation is treated within the generalised gradient approximation (GGA) using Perdew-Burke-Ernzerhof (PBE) functional \cite{JPP6}. Since it is well known that the XANES spectra maps the unoccupied density of states of a material, we display in Fig. \ref{THE}(a) atom projected partial density of states of the W (5d). The form of density of states as a function of energy would predict an overall nature of the XANES spectra and their formal similarity is clearly visible in Fig. \ref{THE} (b).  In Fig. \ref{THE}(b) we present the simulated L$_3$ absorption spectra of pure bulk crystalline bcc W for two different core-level broadening ($\tau$), 10.0 eV (black solid line) and 15.0 eV (red solid line) respectively. Clear resemblance between the theoretically calculated XANES of pure bulk W crystal and that of experimentally observed spectrum presented in Fig. \ref{XANES} confirms our hypothesis (that is, there is no significant signature of change of chemical state of W).  Theoretical spectra with $\tau$ = 15.0 eV (red solid line) matches nicely with that of the experimental results. This suggests a larger core level broadening of W.\\
\hspace*{6pt}
We have also carried out grazing incidence x-ray diffraction (GIXRD) measurements for the W-B$_4$C multilayer at 15.5 keV x-ray energy to determine formation of various crystalline phases (if any) inside the multilayer medium. Figure \ref{GIXRDML} shows the measured GIXRD pattern of the W-B$_4C$ multilayer consisting of N=15 bilayer repetitions at a fixed grazing incidence angle of 1$^o$. It may be noted here that the attenuation length of the incident x-ray at 15.5 keV is greater than the total thickness of the W-B$_4$C multilayer stack. The GIXRD measurements show two broad peaks corresponding to the W(110) and W(211) planes, which suggest that W is present in the polycrystalline phase with a very small grain size boundary. We could also observe a very small peak at $\sim$29.9$^o$ that arises due to asymmetric reflection from the Si substrate. The GIXRD measurement does not provide any evidence of formation of any other chemical phases of the W inside the W-B$_4$C multilayer structure.\\
\hspace*{6pt}
In practice, it is very difficult to map depth resolved chemical information on thin layered materials by using conventional x-ray absorption fine structure (XAFS) measurements (transmission or fluorescence mode). The combined XSW-XAFS analysis approach offers an opportunity to study depth resolved local structure of a thin layered materials with atomic scale resolution.
\section{Conclusions}
\hspace*{6pt}
We have demonstrated that the XANES characterization along with the combined XRR-GIXRF analysis can be used as a sensitive probe to perform depth resolved chemical speciation of a thin film structure. We have applied this approach to investigate detailed microstructural properties of the W-B$_4$C superlattice structures. Two sets of W-B$_4$C multilayer structures consisting of N=15 and N=10 bilayer periods have been analyzed using Indus-2 synchrotron and Elettra Sincrotrone Trieste facilities. XSW-XANES results reveal significant changes in the chemical nature of W atoms that are present at the surface and interfaces as compared those present in the bulk thin film medium. This is mainly due to the surface electronic states of the W atoms.   On the other hand, the electronic behaviour of the W atoms that are present deep inside the bulk W layer remains unchanged, which represents the bulk electronic states of W. This is further confirmed through the first principles calculations, carried out on bulk crystals of W. The grazing incidence x-ray diffraction results also support the above mentioned conclusions. The understanding of distribution of surface states as a function of depth for thin films is often necessary to correlate electronic, magnetic and transport properties of a thin film device.\\
\section{Acknowledgements}
\hspace*{12pt}
The authors  acknowledge Sh. M. N. Singh, Indus Synchrotrons Utilisation Division for his help in GIXRD measurements. The authors are deeply indebted to Elettra Sincrotrone Trieste for successful beamtime under IAEA beamtime quota. One of the authors Gangadhar Das would like to thank Homi Bhabha National Institute, India for providing research fellowship.\\
$^*$\textcolor{blue}{mktiwari@rrcat.gov.in}

\begin{thebibliography}{}
\bibitem{JC7} J. Chakhalian, J. W. Freeland, H.U. Habermeier, G. Cristiani, G. Khaliullin, M. van Veenendaa, and B. Keimer, Science  \textbf{318}, 1114-1117 (2007).
\bibitem{NN6} N. Nakagawa, H. Y. Hwang, and D. A. Muller, Nature materials \textbf{5}, 204-209 (2006).
\bibitem{EB1} E. Benckiser et al., Nature materials, \textbf{10},189-193 (2011).
\bibitem{ADR8} A. D. Rata, A. Herklotz, K. Nenkov, L. Schultz, and K. Dörr,  Phys. Rev. Lett. \textbf{100}, 076401 (2008).
\bibitem{GJ8}  G. Jackeli and G. Khaliullin, Phys. Rev. Lett. \textbf{101}, 216804 (2008).
\bibitem{AT8}  A. Tebano, C. Aruta, S. Sanna, P. G. Medaglia, G. Balestrino, A. A. Sidorenko, R. De Renzi, G. Ghiringhelli, L. Braicovich, V. Bisogni, and N. B. Brookes, Phys. Rev. Lett. \textbf{100}, 137401 (2008).
\bibitem{CA9}  C. Aruta, G. Ghiringhelli, V. Bisogni, L. Braicovich, N. B. Brookes, A. Tebano, and G. Balestrino, Phys. Rev. B \textbf{80}, 014431 (2009).
\bibitem{MS9}  M. Salluzzo, J. C. Cezar, N. B. Brookes, V. Bisogni, G. M. De Luca, C. Richter, S. Thiel, J. Mannhart, M. Huijben, A. Brinkman, G. Rijnders, and G. Ghiringhelli, Phys. Rev. Lett. \textbf{102}, 166804 (2009).
\bibitem{SB7} S. Bera, K. Bhattacharjee, G. Kuri, and B. N. Dev,  Phys. Rev. Lett. \textbf{98}, 196103 (2007); P. Yu, J. S. Lee, S. Okamoto, M. D. Rossell, M. Huijben, C. H. Yang, Q. He, J. X. Zhang, S. Y. Yang, M. J. Lee, Q. M. Ramasse, R. Erni, Y. H. Chu, D. A. Arena, C. C. Kao, L. W. Martin, and R. Ramesh,  Phys. Rev. Lett. \textbf{105}, 027201 (2010).
\bibitem{MZ5} M. Zwiebler, J. E. Hamann-Borrero, M. Vafaee, P. Komissinskiy, S. Macke, R. Sutarto, F. He, B. Büchner, G. A. Sawatzky, L. Alff and J. Geck, New J. Phys. \textbf{17}, 083046 (2015); A. L. Dadlani, S. Acharya, O. Trejo, F. B. Prinz, and J. Torgersen, ACS Appl. Mater. Interfaces, \textbf{8} , 14323-14327 (2016).
\bibitem{JM0} J. Mannhart, and D. G. Schlom, Science \textbf{327} (2010); B. Revaz, M. C. Cyrille, B. L. Zink, I. K. Schuller, and F. Hellman, Phys. Rev. B \textbf{65}, 094417 (2002).
\bibitem{TO1} T. Okumura, T. Nakatsutsumi, T. Ina,Y. Orikasa, H. Arai, T. Fukutsuka, Y.  Iriyama, T.  Uruga, H. Tanida, Y. Uchimoto, and Z. Ogumi, J. Mater. Chem. \textbf{21}, 10051-10060 (2011); C. Jia, Q. Liu, C. Sun, F. Yang, Y. Ren, S. M. Heald, Y. Liu, Z. Li, W. Lu, and J. Xie, ACS Appl. Mater. Interfaces \textbf{6}, 17920-17925 (2014).
\bibitem{GAPL6} G. Das, A. Khooha, A. K.Singh, A. K.Srivastava, and M. K.Tiwari, Appl. Phys. Lett. \textbf{108}, 263109 (2016)
\bibitem{JRC5}J. R. Church, C. Weiland, and R. L. Opila, Appl. Phys. Lett. 106, 171601 (2015); R. Unterumsberger, B.  Pollakowski, M.  Muller, and B.  Beckhoff, Anal. Chem. {\bf 83}, 8623 (2011).
\bibitem{MPRB9} M. K. Tiwari, K. J. S. Sawhney, Tien-Lin Lee, S. G. Alcock, and G. S. Lodha, Phys. Rev. B {\bf 80}, 035434 (2009); M. K. Tiwari, H. Wang,
K. J. S. Sawhney, M. Nayak, and G. S. Lodha, Phys. Rev. B \textbf{87}, 235401 (2013); J. D. Emery, B. Detlefs, H. J. Karmel, L. O. Nyakiti, D. K. Gaskill, M. C. Hersam, J. Zegenhagen, and M. J. Bedzyk, Phys Rev Lett. {\bf 111}, 215501 (2013); J. A. Libera, H. Cheng, M. Olvera de la Cruz, and M.  J. Bedzyk, J. Phys. Chem. B  \textbf{109}, 23001-23007(2005).
\bibitem{JZ3} J. Zegenhagen and A. Kazimirov, {\it The X-Ray Standing Wave Technique: Principle and Applications},World Scientific-(2013), ISBN: 978-981-277-900-7. 
\bibitem{YK5} Y.  Kayser, J. Sa, and J. Szlachetko, Anal. Chem. \textbf{87}, 10815-10821 (2015).
\bibitem{DCM0} D. C. Meyer, K. Richter, P. paufler, P. Gawlitza, and T. Holz, J. Appl. Phys. \textbf{87}, 7218 -7226 (2000).
\bibitem{DCM9} D. C. Meyer, P. Gawlitza, K. Richter, and P. Paufler, J. Phys. D: Appl. Phys. \textbf{32}, 3135-3139 (1999).
\bibitem{BP8} B. Pollakowski, B. Beckhoff, F. Reinhardt, S. Braun, and P.r Gawlitza, Phys. Rev. B \textbf{77}, 235408 (2008); M. Pagels, F. Reinhardt, B. Pollakowski, M. Roczen, C. Becker, K. Lips, B. Rech, B. Kanngießer, and B. Beckhoff, Nuclear Instruments and Methods in Physics Research B \textbf{268}, 370-373 (2010).
\bibitem{BP3} B. Pollakowski, P. Hoffmann, M. Kosinova, O. Baake, V. Trunova, R. Unterumsberger, W. Ensinger, and B. Beckhoff, Anal. Chem. \textbf{85}, 193, (2013).
\bibitem{KS7} K. Sanyal,A. Khooha,G. Das,M. K. Tiwari,and N. L. Misra, Anal. Chem. \textbf{ 89} , 871-876 (2017).
\bibitem{EO6} E. Oakton, G. Siddiqi, A. Fedorov, and C. Coperet, New J. Chem. \textbf{40}, 217-222( 2016).
\bibitem{UJ4} U. Jayarathne, P. Chandrasekaran, A. F. Greene, J. T. Mague, S. DeBeer, K. M. Lancaster, S. Sproules,  and J. P. Donahue, Inorg. Chem. \textbf{53}, 8230-8241 (2014).
\bibitem{FJG5} F. J. Garcia-Garcia, J. Gil-Rostra, F. Yubero, J. P. Espino s, and A. R. Gonzalez-Elipe, J. Phys. Chem. C \textbf{119}, 644-652 (2015).
\bibitem{VL7} V. Luca, M. G. Blackford, K. S. Finnie, P. J. Evans, M. James, M. J. Lindsay, M. Skyllas-Kazacos, and P. R. F. Barnes, J. Phys. Chem. C  \textbf{111}, 18479-18492(2007); D. E. Clinton, D. A. Tryk, I. T. Bae, F. L. Urbach, M. R. Antonio, and D. A. Scherson, J. Phys. Chem.  \textbf{100}, 18511-18514 (1996); S. Yamazoe, Y. Hitomi, T. Shishido, and T. Tanaka, J. Phys. Chem. C \textbf{ 112}, 6869-6879 (2008).
\bibitem{SAG6} S. A. GAUNS. \textit{X-ray Spectroscopic Study of Tungsten Compounds}, PhD Thesis, Department of Physics-Goa University, India (1996).
\bibitem{DPG} G. S. Lodha, RRCAT News Lett. {\bf 22}, 7 (2009).
\bibitem{GRSI5} G. Das, S. R. Kane, A. Khooha, A. K. Singh, and M. K. Tiwari, Rev. Sci. Instrum. {\bf 86}, 055102 (2015); G. Das, A. Khooha, S. R. Kane, A. K. Singh, M K Tiwari. \textit{AIP Conf. Proc}. \textbf{1728}, 020142 (2016).
\bibitem{GJAAS4} G. Das, M. K. Tiwari, A. K. Singh,  and H. Ghosh, J. Anal. At. Spectrom. \textbf{29}, 2405 (2014).
\bibitem{WL2} W. LI, J. Zhu,  X. Ma,  H. LI, H.   Wang, K. J. S. Sawhney,  and Z. Wang, Rev. Sci. Instrum. \textbf{83}, 053114-4 (2012).
\bibitem{MJSR3}  M. K. Tiwari, P. Gupta, A. K. Sinha, S. R. Kane, A. K. Singh, S. R. Garg, C. K. Garg, G. S. Lodha, and S. K. Deb, J. Synchrotron Radiat. {\bf 20}, 386 (2013).
\bibitem{GS6} G. Singh, A., A. Banerji, K. Barpande, V. Bhatnagar, A. G. Bhujle, A. A. Fakhri, P. Fatnani, A. D. Ghodke, and P.R. Hannurkar, Proceedings of Indian Particle Accelerator Conference 94 (2006); A.  K.  Sinha,  A.  Sagdeo,  P.  Gupta,  A. Upadhyay,  A.  Kumar,  M.  N. Singh, R. K. Gupta, S. R. Kane, A. Verma, and S. K. Deb, Journal of Physics: Conference Series \textbf{425}, 072017 (2013).
\bibitem{MXP6} M. K. Tiwari,  and G. Das, X-Ray Spectrom. \textbf{45}, 212-219 (2016).
\bibitem{JMA4} J. M. André, R. Barchewitz, A. Maquet, and R. Marmoret, Phys. Rev. B \textbf{29}, 6576-6585 (1984).
\bibitem{RS7} R. Soufli and E. M. Gullikson, Appl. Opt. \textbf{36}, 5499-5507 (1997).
\bibitem{PP9} P. Pfalzer, J. P. Urbach, M. Klemm, S. Horn, M. L. denBoer, A. I. Frenkel, and J. P. Kirkland, Phys. Rev. B \textbf{60}, 9335 (1999).
\bibitem{WL5} W. Li, X. Yuan, J. Zhu, J. Zhu, and Z. Wang, Phys. Scr. \textbf{90}, 015804 (2015).
\bibitem{PCP6} P. C. Pradhan, A. Majhi, M. Nayak, M. Nand, P. Rajput, D. K. Shukla, A. Biswas, S. K. Rai, S. N. Jha, D. Bhattacharyya, D. M. Phase,and N. K.  Sahoo, J. Appl. Phys.\textbf{ 120}, 045308 (2016).
\bibitem{PNR3} P. N. Rao, S. K. Rai, M. Nayak, and G. S. Lodha, Applied Optics {\bf 52}, 6126 (2013).
\bibitem{PS0} P. Siffalovic, M. Jergel, L. Chitu, E. Majkova, I. Matko, S. Luby, A. Timmann, S. V. Roth, J. Keckes, G. A. Maier, A. Hembd, F. Hertlein, and J. Wiesmann, J. Appl. Crystallogr. {\bf 43}, 1431 (2010).
\bibitem{SC5} S. Clerck et al. Zeitschrift fuer Kristallographie \textbf{220}, 567-570 (2005).
\bibitem{JPP6} J. P. Perdew, K. Burke, and M. Ernzerhof, Phys. Rev. Lett. \textbf{77}, 3865 (1996).
\end{thebibliography}
\end{document}